\documentclass[12pt,fleqn]{article}
\usepackage[DIV12]{typearea}
\usepackage{amsmath,amsfonts,amssymb}
\usepackage{graphicx}
\usepackage{cite}
\usepackage{mathrsfs}
\usepackage{bbm}
\usepackage{pifont}
\usepackage{units}
\usepackage[small,loose]{subfigure}  
\usepackage{color}
\usepackage{psfrag}
\usepackage{datetime}
\usepackage{collref}

\newcommand{\eVdist}{\kern-0.06em}

\newcommand{\gev}{\:\text{Ge\eVdist V}}
\newcommand{\tev}{\:\text{Te\eVdist V}}

\newcommand{\SU}[1]{\ensuremath{\mathrm{SU}(#1)}}
\newcommand{\U}[1]{\ensuremath{\mathrm{U}(#1)}}

\allowdisplaybreaks[1]

\title{The heterotic string yields natural supersymmetry}

\begin{document}

\begin{titlepage}

\begin{flushright}
TUM--HEP 826/12
\end{flushright}

\vspace*{1.0cm}

\begin{center}
{\Large\bf
The heterotic string yields natural supersymmetry
}

\vspace{1cm}

\textbf{
Sven Krippendorf\footnote[1]{Email: \texttt{krippendorf@th.physik.uni-bonn.de}}{}$^a$,
Hans Peter Nilles\footnote[2]{Email: \texttt{nilles@th.physik.uni-bonn.de}}{}$^a$,
Michael Ratz\footnote[3]{Email: \texttt{michael.ratz@tum.de}}{}$^b$,\\
Martin Wolfgang Winkler\footnote[4]{Email: \texttt{martin.winkler@tum.de}}{}$^b$
}
\\[5mm]
\textit{\small
{}$^a$ Bethe Center for Theoretical Physics\\
{\footnotesize and}\\
Physikalisches Institut der Universit\"at Bonn,\\
Nussallee 12, 53115 Bonn, Germany
}
\\[3mm]
\textit{\small
{}$^b$ Physik--Department T30, Technische Universit\"at M\"unchen, \\
~~James--Franck--Stra\ss e, 85748 Garching, Germany
}
\end{center}

\vspace{1cm}

\begin{abstract}
The most promising MSSM candidates of the heterotic string reveal some
distinctive properties. These include gauge--top unification, a specific
solution to the $\mu$--problem and mirage pattern for the gaugino masses. The
location of the top-- and the Higgs--multiplets in extra dimensions differs
significantly from that of the other quarks and leptons leading to a
characteristic signature of suppressed soft breaking terms, reminiscent of a
scheme known as natural supersymmetry.
\end{abstract}

\end{titlepage}

\newpage

String theory might provide us with a consistent ultraviolet completion of the 
minimal supersymmetric standard model (MSSM) with unified gauge and 
gravitational couplings. To analyze this ansatz we have to identify ways 
to embed the MSSM into string theory and then study properties of 
realistic models. In the present paper we report on progress in model 
building within the heterotic string \cite{0611095,0807.4384,0708.2691} and its implications for 
supersymmetry (SUSY) at the large hadron collider (LHC) at CERN. The emergent 
picture from the heterotic braneworld can be summarized as follows:

\begin{itemize}

\item   large gravitino mass ($m_{3/2}$)  and heavy 
string moduli, all in the multi--TeV range or even heavier
(at least of order $m_{3/2}\cdot\log (M_\mathrm{Planck}/m_{3/2})$),
\item   gaugino masses and $A$--terms in the TeV--range, 
suppressed with respect to $m_{3/2}$  by a factor 
$\log (M_\mathrm{Planck} / m_{3/2}),$

\item  a mirage pattern for gaugino masses (compressed spectrum),

\item    top--squarks $(\tilde t_\mathrm{L}, \tilde b_\mathrm{L})$ and $\tilde
t_\mathrm{R}$ 
in the TeV--range,

\item    other squarks in the multi--TeV--range of order the 
gravitino mass.

\end{itemize}
These properties are similar in some aspects to a bottom--up  approach called
``natural SUSY''.\footnote {The terminology ``natural
SUSY''  appeared to our knowledge first in \cite{0509221},
where references to the earlier  literature on similar models with
non--universal scalar masses can be found. There it was motivated from bottom-up
arguments while here it emerges in a top--down construction  from heterotic
string theory.}
The origin of the pattern can be traced back to two distinct properties  of
realistic MSSM candidates from heterotic string theory:  (i) specific
localization properties of fields (specifically the top quark)  in extra
dimensions and (ii) the appearance of mirage mediation (mixed modulus anomaly
mediation) of supersymmetry breakdown. This mirage
pattern~\cite{0702146}\footnote {For an explanation of the terminology 
see~\cite{0509158}.} seems to be pretty generic in string theory. It was first
observed~\cite{0411066,0503216,0504036} in type IIB theory in the framework of
the KKLT scenario~\cite{0301240} and it appears naturally in the heterotic
string  theory as well~\cite{0802.1137}. It is characterized by the appearance
of a factor 
\begin{equation}
 \log (M_\mathrm{Planck}/ m_{3/2} )
\end{equation}
that suppresses the soft terms of modulus mediation compared to the gravitino 
mass (and enhances the masses of the moduli by the same factor).  Radiative
corrections to the soft terms as in anomaly
mediation~\cite{9810155,9810442,9911029}  become competitive resulting in a
mixed modulus--anomaly mediation.  Specific properties of the MSSM
$\beta$--functions (negative for $\SU3$, positive for
$\SU2$ and $\U1$) lead to the appearance
of a mirage scale,  where soft terms coincide. This leads to a compressed
spectrum of  soft terms that improves the so--called little hierarchy
problem~\cite{0511320},  improves precision gauge coupling
unification~\cite{0911.4249} and alters predictions  --- compared to pure modulus
mediation --- for potential LHC observation  significantly, see~\cite{1110.1287}
and references therein.  Properties of the mirage scheme turn out to be pretty
robust for  gaugino masses and $A$--parameters, while soft scalar masses are 
strongly model dependent. In fact, it was shown in~\cite{0603047} that  masses
of squarks and sleptons are less protected and tend to become  as large as the
gravitino mass. This scheme, with sfermion masses in  the multi--TeV--range and
gaugino masses (and $A$--terms) at the  TeV--scale seems to be pretty generic in
Type II and heterotic string theory.

A more detailed picture requires explicit model building and this  brings us to
the main result of this paper. Such a picture arises from model  building in
heterotic string theory as observed e.g.\ in the 
Minilandscape~\cite{0611095,0807.4384,0708.2691} of orbifold compactifications.
The benchmark  models presented there~\cite{0708.2691} show distinctive
properties shared by  a majority of the models. There is only one pair of Higgs
doublets (no Higgs triplets), $H_u$ and $H_d$, both in the untwisted sector
such that the Higgs bilinear $H_u\,H_d$ is neutral under all
selection rules. This leads to a solution of  the $\mu$--problem via a
(discrete) $R$--symmetry~\cite{0812.2120,1003.0084} and guarantees  Minkowski vacua before supersymmetry
breakdown. In these realistic models,  the top quark plays a special role. Both
$(t_\mathrm{L}, b_\mathrm{L})$ and $t_\mathrm{R}$ ``live'' in the  untwisted
sector while other quark-- and lepton--multiplets reside in  various twisted
sectors. As a result we have only one non--vanishing  Yukawa coupling at the
trilinear level consistent with gauge--top--Yukawa
unification~\cite{Faraggi:1991be,0905.3323}.\footnote{Other Yukawa couplings are
suppressed as in the framework of the
Frogatt--Nielsen mechanism \cite{Froggatt:1978nt}.} This is a direct consequence
of the fact that both the Higgs multiplets and the top multiplets ``live'' in
the bulk (untwisted sector), while other particles are localised at fixed points
or fixed tori in the extra dimensions. As we shall see, this particular
configuration has important consequences for  the soft mass terms of $(\tilde
t_\mathrm{L}, \tilde b_\mathrm{L})$-- and $\tilde t_\mathrm{R}$--squarks as  well
as for the soft Higgs masses. Fields in the untwisted sector descend  from a
torus compactification of extra dimensions. Torus compactification  in itself
would yield $N=4$ supersymmetry in $d=4$ and the untwisted sector  of orbifold
compactification feels remnants of this extended supersymmetry,  most clearly
seen in the framework of ``no--scale'' models~\cite{Cremmer:1983bf}. In the
models  under consideration this gives a suppression to the soft masses of 
$(\tilde t_\mathrm{L}, \tilde b_\mathrm{L})$ and $\tilde t_\mathrm{R}$ not
shared by the others squarks  and sleptons. We thus obtain the pattern of soft
terms with a two step  hierarchy: gauginos, Higgses and stops at the TeV scale,
all other sfermions  at the multi--TeV scale of the order of the gravitino mass.
A large  Yukawa coupling for the top quark requires special geometric
properties  of extra dimensions which reflect themselves in the pattern of soft
scalar  masses. In upshot, the soft masses of the top--multiplet are so light
because the  mass of the top quark is so large.

Let us now discuss the mechanism of SUSY breakdown in more detail.  The models
of the Minilandscape show a specific pattern of  gauge group in the hidden
sector~\cite{0611203} with SUSY breakdown via gaugino 
condensates~\cite{Nilles:1982ik,Nilles:1982xx}. Moduli stabilization can proceed
along the  lines of~\cite{1012.4574,1102.0011}. This could take care of the $U$--
and $T$--moduli of the models,  but not the dilaton $S$. We thus remain with a
``run--away''  dilaton and a positive ``vacuum energy'' for finite $S$. The
vacuum  energy has to be adjusted to zero. This can be done with a scalar
matter  field $X$ (in the untwisted sector) in a ``down--lifting mechanism'' as 
described in~\cite{0802.1137}. This adjusts the vacuum energy and fixes the
vacuum expectation value of the dilaton $S$. A mirage picture of  mixed
dilaton--anomaly mediation emerges.

Such settings in which supersymmetry is dominantly broken by matter fields have been
studied in \cite{0603047}. The relevant K\"ahler potential reads
\begin{equation}\label{eq:Kaehler1}
 K~=~ -3 \ln \left(T+ \overline{T}\right) 
 + X\, \overline{X} 
 + Q_\alpha \overline{Q}_\alpha\, (T+ \overline{T})^{n_\alpha} \Bigl[
 1+ \xi_\alpha\, X\, \overline{X} + \mathcal{O} (|X|^4)
\Bigr] \;,
\end{equation}
where $X$ denotes a ``hidden'' matter field the $Q_\alpha$ are the observable
fields with  ``modular weights'' $n_\alpha$. The general formulae~\cite{9303040,9707209} for the soft masses have been
specialized to the case that supersymmetry is dominantly broken by $X$ with $F^X\neq 0$ and $\langle X\rangle\approx 0$~\cite{0603047}.
Following~\cite{0802.1137}, we define the quantity
\begin{equation}
 \varrho~:=~\frac{16 \pi^2}{m_{3/2}}\frac{F^S}{S_0+\overline{S_0}}\;,
\end{equation}
where $S_0\in\mathbbm{R}$ is the VEV of the dilaton. The soft supersymmetry breaking parameters are then \cite{0802.1137}
\begin{subequations}\label{eq:LNformulae}
\begin{eqnarray}
 M_a & = & \frac{m_{3/2}}{16 \pi^2}\,\left[\varrho+b_a\,g_a^2\right]\;,\\
 A_{\alpha\beta\delta} & = & 
 \frac{m_{3/2}}{16 \pi^2}\,\left[-\varrho+\left(\gamma_\alpha+\gamma_\beta+\gamma_\delta\right)\right]
 \;,\\
 m_\alpha^2 & = & \frac{m_{3/2}^2}{(16\pi^2)^2}\,\left[\varrho^2\xi_\alpha-\Dot{\gamma}_\alpha
 	+2\varrho\,\left(S_0+\overline{S}_0\right)\,\partial_S\gamma_\alpha
	+(1-3\xi_\alpha)(16\pi^2)^2\right]\;.
\end{eqnarray}
\end{subequations}
Here $b_a$ denote the usual MSSM $\beta$--function coefficients, $\gamma_\alpha$
the standard anomalous dimensions and the derivative of $\gamma_i$ with respect
to $S$ is given by $( S_0+\overline{S}_0 )\,\partial_S \gamma_i =  -
\gamma_i$.\footnote{We assume that the holomorphic Yukawa couplings do not
depend on the dilaton.}

An important feature of the scalar masses $m_{\alpha}$ is that they are
generically of the order $m_{3/2},$ unless $\xi_\alpha=1/3$. Untwisted matter
fields $Q_{\alpha}^{\rm ut}$ can lead to a situation with $\xi_\alpha=1/3.$ Of
course, in specific models we do not expect that $\xi_\alpha$ for untwisted
matter fields, denoted by $\xi_3$ from now on, is exactly equal to $1/3,$ but we
still expect that generically untwisted sector fields will have a value of
$\xi_\alpha$ closer to $1/3$ than twisted matter fields, resulting in a
hierarchy between the respective soft mass terms. In this study we base our
discussion on the K\"ahler potential with one so--called overall K\"ahler
modulus $T$ (cf.~equation 11 in~\cite{Witten:1985xb})
\begin{equation}\label{eq:Knaive}
 K~=~-3\ln\left[T+\overline{T}
 -\frac{1}{3}\left(Q_\alpha^\mathrm{ut}\,\overline{Q}_\alpha^\mathrm{ut}
 +\widetilde{X}\,\overline{\widetilde{X}}\right)\right]\;,
\end{equation}
where $\widetilde{X}$ is the not yet canonically normalized field breaking
supersymmetry. This K\"ahler potential describes untwisted fields
$Q_\alpha^\mathrm{ut}$ and $\widetilde{X},$ but not twisted sector fields for which a
different K\"ahler potential is required. We can bring the above K\"ahler
potential to the form in equation~\eqref{eq:Kaehler1}, 
\begin{equation}
 K~=~-3 \ln (T+ \overline{T}) + X\, \overline{X} 
 + \frac{Q^\mathrm{ut}_\alpha \overline{Q}^\mathrm{ut}_\alpha}{(T+ \overline{T})}\, \Bigl[
 1+\frac{1}{3}\, X\, \overline{X} + \mathcal{O} (|X|^4)
\Bigr] \;,
\end{equation}
where we went to canonically normalized hidden sector matter fields
$(X=\widetilde{X}/(T+\bar{T})^{1/2})$. If we assume that supersymmetry is
dominantly broken by an $F$--term VEV of $X$, we get from \eqref{eq:LNformulae}
soft masses for $Q_\alpha^\mathrm{ut}$ which are highly suppressed against
$m_{3/2}$.\footnote{Observe that it is important that  the field $X$ itself is
an untwisted sector field.}  As mentioned above, for twisted sector matter
fields the K\"ahler potential is different from \eqref{eq:Knaive} and one
obtains $\xi_\alpha$ values, denoted by $\xi_f$ from now onwards, for such
fields which differ from $1/3$. In particular, given the K\"ahler potential
\eqref{eq:Knaive}, we see that
\begin{equation}
 m_\alpha~\sim~\left\{\begin{array}{ll}
  \displaystyle \frac{m_{3/2}}{16\pi^2}\; \quad & \text{for $Q_\alpha^{\mathrm{ut}}$} \;,\\
  m_{3/2}\;\quad & \text{otherwise} \;.
 \end{array}\right.
\end{equation}
As mentioned before, in the explicit heterotic string models
$(t_\mathrm{L},b_\mathrm{L})$, $t_\mathrm{R}$ and $H_{u,d}$ are in the untwisted
sector while the other MSSM fields are not. Taking into account that the gaugino
masses are suppressed against $m_{3/2}$ by a factor
$\log(M_\mathrm{Planck}/m_{3/2})$, we then obtain the ``natural SUSY'' pattern
from explicit model building in heterotic string theory. This pattern of
soft masses is markedly different from the spectra in the CMSSM and leads to a
specific pattern that can be tested at the LHC.

What are the specific properties of such a scheme? The main challenges come from
a discussion of potential tachyonic instabilities. Remember that we are
discussing a unified model originating  from string theory at a very high scale
($M_\mathrm{GUT} \sim 10^{16}\,$GeV) and we have to analyze the running of mass
parameters from this high scale to the TeV scale. This is different from
bottom--up approaches where we just assume a consistent spectrum at the TeV
scale. We perform the running of mass parameters using the spectrum
generator \texttt{Softsusy}~\cite{0104145}. In our analysis we will  require the
absence of tachyons at all scales.\footnote{We are aware that  this assumption
could be too strict~\cite{0806.3648}, but we stick to it here for simplicity.} In
particular the small value of the stop mass is a potential source of
instability since we obtain at the GUT--scale\footnote{This relation is sensitive to the Yukawa couplings at the GUT scale. It slightly changes for
large $\tan\beta$ where $y_b$ becomes sizeable.}

\begin{equation}
 m_{Q_{3_\mathrm{L}}}^2(M_{\text{GUT}}) ~\simeq~ \frac{m_{3/2}^2}{(16 \pi^2)^2} \left( \xi_3\,\varrho^2 - 3.7\,\varrho+ 0.8 +(16\pi^2)^2(1-3\xi_3)\right) .
 \label{eq:tachyonicq3l}
\end{equation}
For $\xi_3=1/3$ the absence of a tachyonic stop/sbottom mass at the GUT--scale requires
$\varrho\gtrsim 10.9$ which corresponds to a mirage scale
\begin{equation}
 M_\mathrm{MIR}~=~M_\mathrm{GUT}\,\mathrm{e}^{-8\pi^2/\varrho} 
 ~\gtrsim 10^{13}\,\gev\;.
\end{equation}
However, any small correction to $\xi_3=1/3$ allows us to avoid tachyonic
boundary conditions even for small $\varrho$. Such corrections may originate for
example from small mixing of the bulk families with localized states or might be
induced by sub--dominant mediation of supersymmetry breaking. In this case a
lower mirage scale can be realized.

Besides the potential tachyonic boundary conditions, the large hierarchy among
the scalar masses can lead to tachyonic stop/sbottom masses as a result of
renormalization group running.  In particular there exists a 2--loop
contribution to the $\beta$--function of $m_{Q_{3\mathrm{L}}}^2$ which reads (see
for example~\cite{9311340})
\begin{equation}\label{eq:twoloop}
 \beta^\mathrm{2-loop} ~\simeq~ \frac{1}{(16\,\pi^2)^2}  \;48 \,g_3^4\,m_f^2\;,
\end{equation}
where $m_f^2=(1-3\,\xi_f)\,m_{3/2}^2$ denotes the common mass of the scalar fields which are not in the untwisted sector.\footnote{Here we neglect subleading contributions from anomaly mediation.} 
Including the 1--loop term $\propto m_{\widetilde{g}}^2$ we obtain
\begin{equation}
 m_{Q_{3\mathrm{L}}}^2(M_Z^2)  ~\sim~ 
 m_{Q_{3\mathrm{L}}}^2 + 5 m_{\widetilde{g}}^2 \left(1 - \frac{(0.1\,m_f)^2}{m_{\widetilde{g}}^2}\right)\;,
\end{equation}
where the quantities on the right--hand side are to be evaluated at the GUT
scale. This suggests that the absence of tachyonic stops/sbottoms at the low
scale requires $m_f \lesssim 10\, m_{\widetilde{g}}$ at the GUT scale.
It turns out that the actual bound is slightly stronger as there exist sub--leading 1--loop terms $\propto m_f^2$
which tend to decrease $m_{Q_{3\mathrm{L}}}^2$ further. The latter are, however, suppressed by small couplings and
small numerical coefficients.

\begin{figure}[h!]
\begin{center}
  \includegraphics[width=10.5cm]{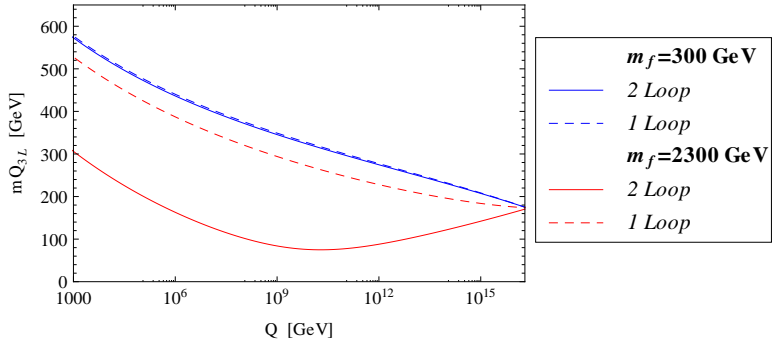}
\end{center}
\caption{RGE running of $m_{Q_{3\mathrm{L}}}$ at the 1--loop and 2---loop level.}
\label{fig:running}
\end{figure}

In figure~\ref{fig:running} we show the full RGE running of $m_{Q_{3\mathrm{L}}}$
including sub--leading terms. As boundaries we have chosen a universal gaugino mass
$m_{1/2}=300\gev$, $A=-m_{1/2}$ and $m_i^2=m_{1/2}^2/3$ for the untwisted
fields. These conditions can be obtained from~\eqref{eq:LNformulae} in the limit
of large $\varrho$ and $\xi_3=1/3$.
It can be seen that 2--loop effects are negligible for small $m_f$. However, with
increasing $m_f$ the term~\eqref{eq:twoloop} becomes more important and
eventually $Q_{3\mathrm{L}}$ becomes tachyonic. 

\vskip0.3cm

Nevertheless we find in parameter scans of the heterotic ``natural SUSY" scheme large regions of parameter space consistent with the following constraints:
\begin{itemize}
 \item no tachyons (at all scales),
 \item correct electroweak symmetry breaking (EWSB),
 \item $115.5\gev<m_h < 127\gev$ (combined LHC and LEP bound
  on the Higgs mass~\cite{ATLASHiggs,CMSHiggs}),
 \item LHC limits on the superpartner mass spectrum,
 \item no colored LSP.
\end{itemize}
In the considered scheme the superpartners of the first two generations
become heavy. LHC constraints on gluinos, stops and sbottoms mainly arise from
searches for jets + missing energy as well as searches for di--lepton signals
(see for example~\cite{1010.0692,1110.6443,1110.6670,1111.2830,1110.6926}). Here
--- based on~\cite{1110.6926} --- we will use the following estimates of the
constraints
\begin{equation}
 m_{\widetilde{t}_1},\,m_{\widetilde{b}_1} ~>~ 250 \gev\;, \qquad
 m_{\widetilde{g}} ~>~ 700 \gev\;.
\end{equation}
If gluinos and stop/sbottom are light the limits become slightly stronger, we
assume
\begin{equation}
 m_{\widetilde{t}_1},\,m_{\widetilde{b}_1} ~>~ 250 \gev + 0.5\,(1000\gev- m_{\widetilde{g}}) 
 \quad \text{for}~ m_{\widetilde{g}} ~=~ 700-1000\gev\;.
\end{equation}
This simple treatment is sufficient for our purposes as we will use the current
LHC sensitivity only for illustration.

\begin{figure}[h!] 
  \begin{minipage}[c]{0.5\textwidth}
    \includegraphics[width=6.7cm]{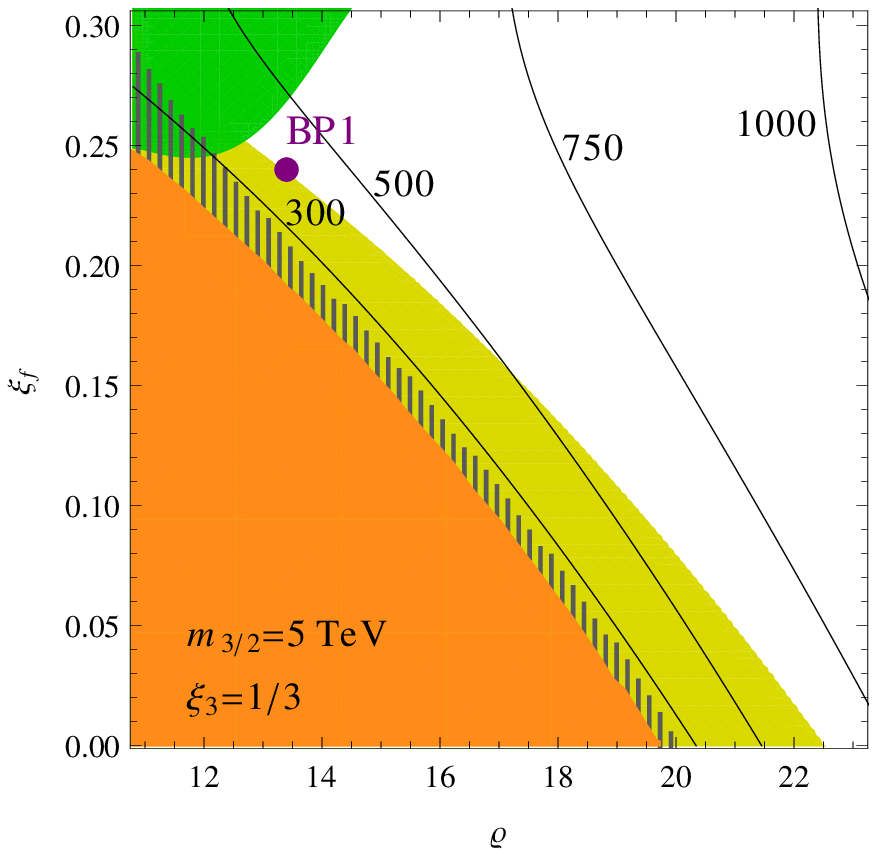}
  \end{minipage}
  \begin{minipage}[c]{0.5\textwidth}
    \includegraphics[width=6.7cm]{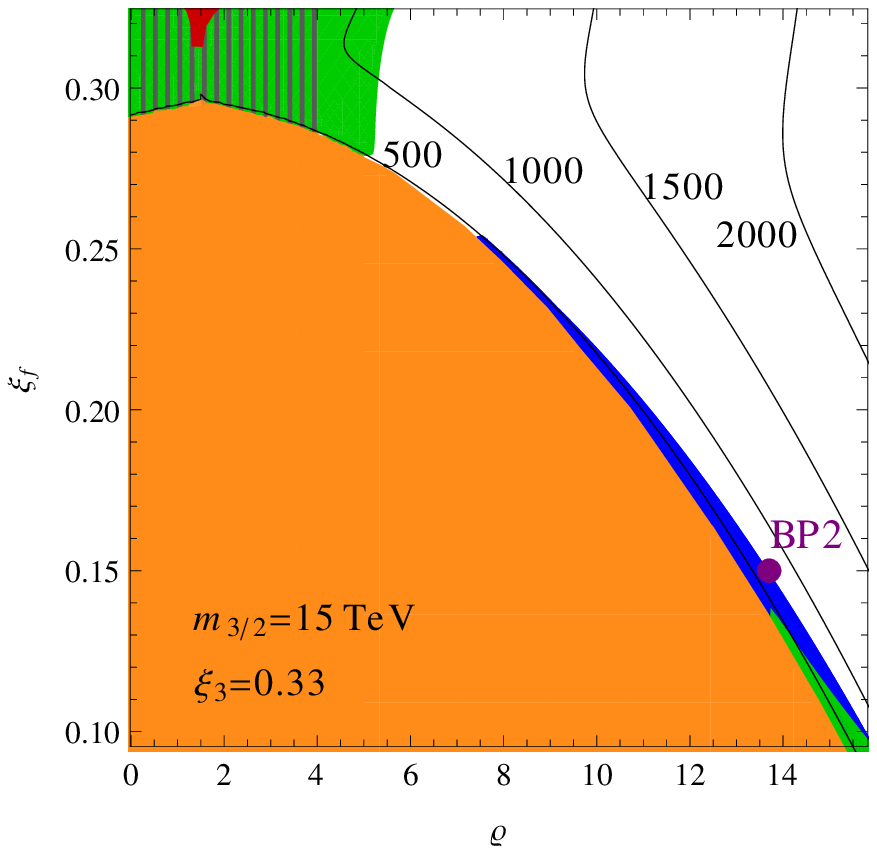} 
  \end{minipage}\\[-3mm]
 \begin{center}
  \includegraphics[width=4.8cm]{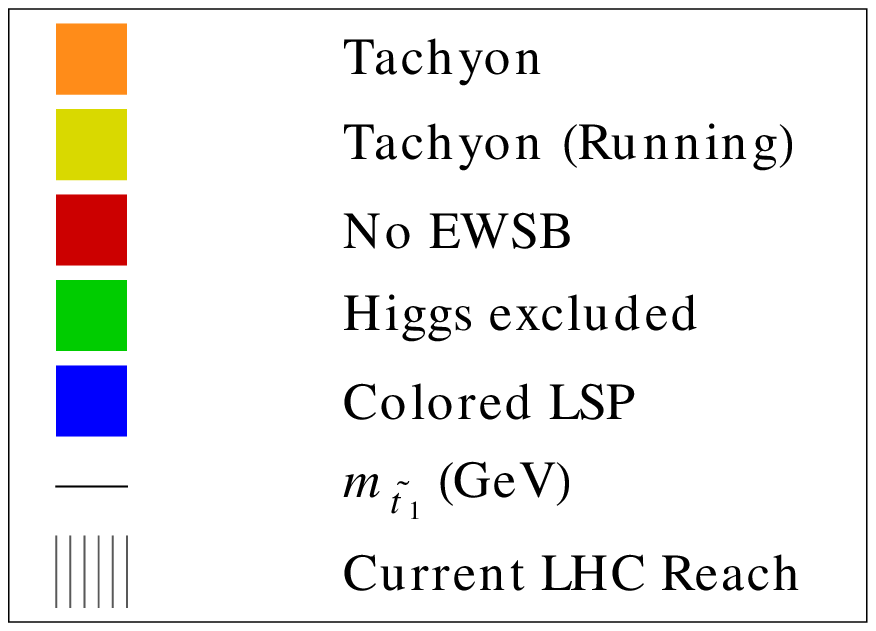} 
\end{center}
\caption{Parameter scans with different gravitino mass. On the left we assume that $\xi_3=1/3$ is exact while on the right we assume that there are corrections giving $\xi_3=0.33$. The colored regions are excluded while the hatched regions indicate the current reach of the LHC (see text). The contours refer to $m_{\widetilde{t}_1}$. The particle spectrum for the two benchmark points BP1 and BP2 is shown in figure~\ref{fig:spec1}.
}
\label{fig:5tev}
\end{figure}

In figure~\ref{fig:5tev} we present two scans in the $\varrho$--$\xi_f$--plane
for fixed gravitino mass and $\xi_3$.\footnote{In the scans we have set
$\text{sgn}\mu= +$ and $\tan\beta=10$.} Requiring the absence of tachyons
substantially constraints the parameter space. For the case where $\xi_3=1/3$ is
exact, the region with $\varrho< 10.9$ is not shown as it generally yields a
negative $m^2_{Q_{3\mathrm{L}}}$ at the GUT scale. Small values of $\varrho$
can, however, be accessed for $\xi_3\neq 1/3$ as shown for the case
$\xi_3=0.33.$ For low values of $\xi_f$, the hierarchy in the scalar sector
grows which tends to decrease $m^2_{Q_{3\mathrm{L}}}$ through the RGE running.
Both scans exhibit a sizeable region (orange) where this effect is so strong
that tachyonic stops/sbottoms are obtained at the weak scale. In the yellow
region $m^2_{Q_{3\mathrm{L}}}$ becomes negative at an intermediate scale, but
--- towards the low scale --- turns positive again. As we require the absence of
tachyons at all scales we also exclude this part of the parameter space.  In the
green region to the left of both scans, the Higgs boson mass is below its
current limit of $115.5\gev$. In the scan to the right there exists also some
parameter space with $m_h>127\gev$ in the lower right corner. As can be seen,
the current LHC searches for superpartners have not yet reached the sensitivity
to constrain the parameters in this scheme further.

\begin{figure}[h!]
\label{fig:3tev}
\begin{center}
 \includegraphics[width=6cm]{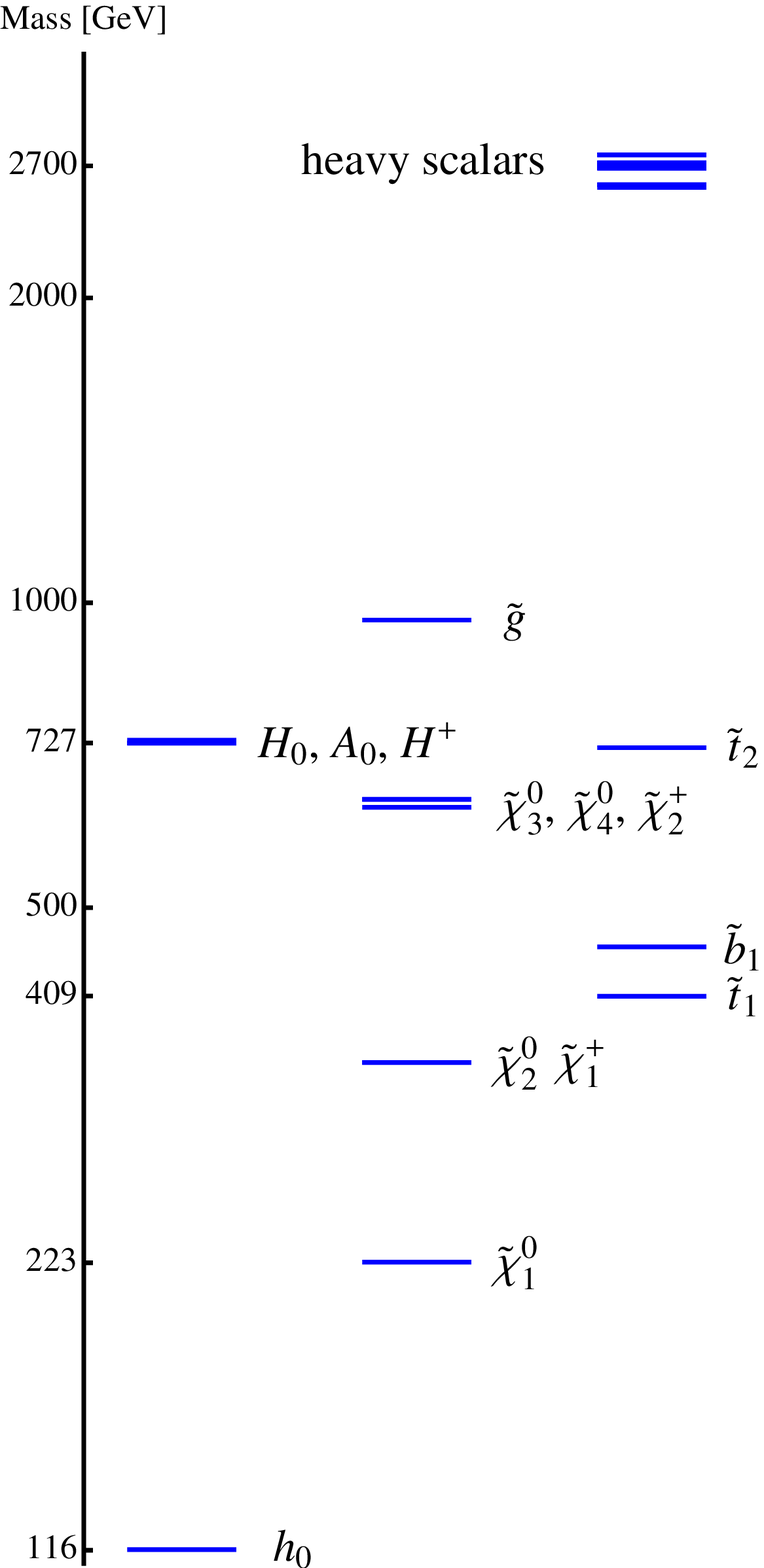}\hspace{1mm}
 \includegraphics[width=6cm]{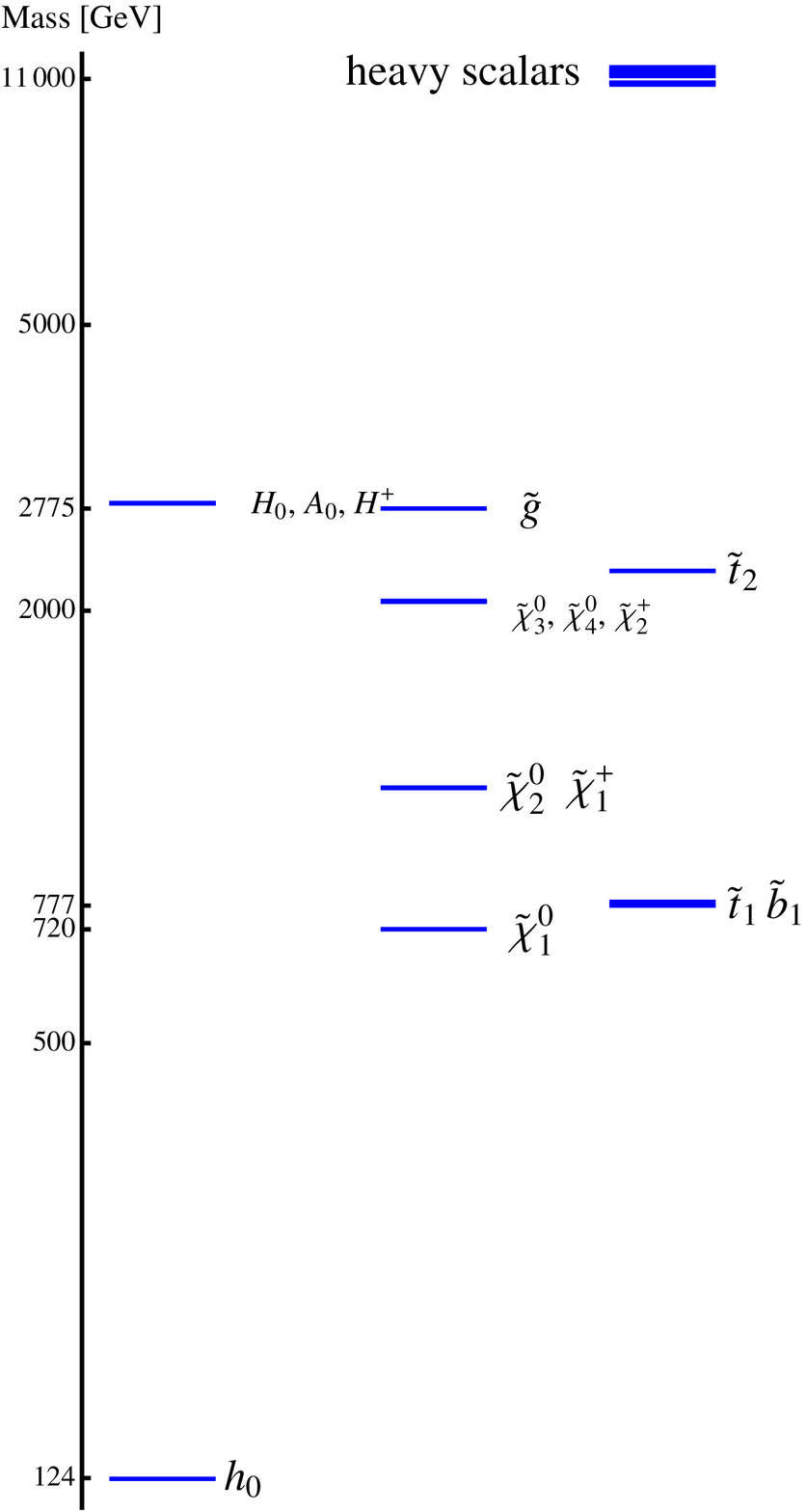}\\
\end{center}
\caption{Particle spectra for the benchmark points BP1 (left) and BP2 (right).
}
\label{fig:spec1}
\end{figure}

The particle spectra for the two benchmark points indicated in
figure~\ref{fig:5tev} are visualised in figure~\ref{fig:spec1}. In both spectra
there is a clear hierarchy: the gauginos, higgsinos and the scalars of the
untwisted sector are significantly lighter than the other superpartners. The
lightest scalars are $\widetilde{t}_1$ and $\widetilde{b}_1$ as their mass is
decreased by the heavy scalars through the RGE running. Due to the mirage
mediation the pattern of gaugino masses is compressed compared to the CMSSM. In
both spectra the lightest superpartner is the bino.\footnote{The relic density
of the binos can potentially match the dark matter density if there are stop
co--annihilations.} The higgsinos are heavier as correct electroweak symmetry
breaking requires $|\mu|\sim |m_{H_u}|$ at the weak scale and $m_{H_u}$ receives
a contribution $\mathcal{O}(m_{\widetilde{g}})$ from the RGE running. As the
scale of supersymmetry breaking is unknown the overall scale of the spectrum
cannot be determined. 

Turning to the Higgs sector, we find that spectra with $m_{\widetilde{t}},m_{\widetilde{g}} \lesssim 1\tev$ yield $m_h < 120 \gev$. Therefore, if the recent hints for $m_h\sim 125\gev$ observed by ATLAS and CMS~\cite{ATLASHiggs,CMSHiggs} are confirmed, this may suggest heavier stops and gluinos.

We see that the heterotic string as a UV completion of the MSSM leads to a
pattern of soft terms that is compatible with all phenomenological constraints.
The most relevant restrictions of the parameter space arise from the potential
appearance of tachyonic instabilities. In fact, this is a situation to be faced
in all top--down constructions where the stop masses are smaller than the masses
of other squarks and sleptons. In the present paper we have considered the case
where tachyonic  instabilities are absent from the large (GUT) scale to the weak
scale. This might be too strong an assumption. Strictly  speaking we would need
this absence only at the weak scale, but such a situation would require a
careful analysis of the cosmological evolution along the lines of
reference~\cite{0806.3648}. This is  beyond the scope of this paper and will be
subject of future research.

Apart from the heterotic string there are other  constructions such as type II 
strings or M--theory that have been discussed in the present context. In the
Type IIB theory with uplifting a la KKLT~\cite{0301240} one finds a mirage
scheme for gaugino masses~\cite{0503216} and heavy squarks and sleptons
(including stops)~\cite{0603047}. Models based on the large volume
scenario~\cite{0502058} could lead to a variety of patterns of soft breaking
terms~\cite{0906.3297,1003.0388,1011.0999}. A similar situation can be found in
F--theory~\cite{0906.3297}, where gauge mediation has been conjectured to be the
major source of supersymmetry breakdown~\cite{0808.1571,0809.1098}.  Models
based on M--theory~\cite{0606262} lead to a pattern similar to that of type IIB a la
KKLT, with a compressed (mirage like) spectrum of gauginos and (ultra) heavy
sfermions (see~\cite{1112.1059} and references therein). In our spectrum of
soft masses from the heterotic string, this hierarchy between the (heavy) scalar
masses and the gluino mass cannot be arbitrarily large (i.e.~there is no
decoupling limit of the heavy scalars) due to the appearance of tachyonic light
scalar masses for too large hierarchies. This hierarchy among the scalar masses
due to the geometric separation of twisted and untwisted matter fields can be
reduced in models with $\xi_3\neq 1/3,$ corresponding to less no--scale
cancellations for the untwisted matter fields. Experiments at the LHC might be
able to distinguish between the various schemes.

The heterotic pattern described here will be a serious challenge  for SUSY
searches at the LHC because of two reasons. First there is the compressed
pattern of gaugino masses typical for the scheme of mirage mediation. It
significantly reduces the ratio of gluino-- to LSP--mass with important
consequences for the properties of the gluino decay chain. Secondly, because of
the light stops, this  decay chain will predominantly include jets of heavy
particles that are more difficult to identify experimentally. Our benchmark
models in figure~\ref{fig:spec1} show that the present reach of LHC does not
yet  restrict the parameter space.

The pattern has characteristic properties relating various types of soft terms
but unfortunately cannot determine the overall scale of the SUSY breakdown. Here
we have considered two benchmarks with small and large value of $m_{3/2}$,
respectively. We can only  hope that this overall scale is small enough to be
within the reach of the LHC.

The theories considered here are the result of a string theory construction
(including a consistent incorporation of gravity) as a UV completion of the
MSSM. They reveal for the first time an explicit relation between MSSM
constructions and the mechanism of SUSY breakdown and mediation from a top--down
point of view. We see a profound connection between location of the fields in
extra dimensions, the size of Yukawa couplings and the pattern of soft mass
terms. The top--quark plays a very special role in this construction. The sector
including the top--quark and the Higgs--multiplets seem to be protected by a
higher  degree of ($N=4$ extended) supersymmetry in extra
dimensions, with important consequences for the phenomenological prediction of
the scheme. We are eagerly waiting for the LHC to test this picture in the not so
distant future.
\vskip0.15cm

\noindent
\textbf{Acknowledgments}\\
We would like to thank Ben Allanach, Matthew Dolan and Jamie Tattersall for
useful conversations. This work was partially supported by the
SFB--Transregio TR33 ÒThe Dark UniverseÓ (Deutsche Forschungsgemeinschaft), the
European Union 7th network program ÒUnification in the LHC eraÓ
(PITN--GA--2009--237920) and the DFG cluster of excellence Origin and Structure of the
Universe (Deutsche
Forschungsgemeinschaft).

\begin{footnotesize}

\bibliography{heteroticsusybib}
\bibliographystyle{h-physrev}
\end{footnotesize}
\end{document}